# Permanent Magnets Undulator for Terahertz FEL

M.L. Petrosyan[1], L.A. Gabrielyan[1], Yu.R. Nazaryan[1], G.Kh.Tovmasyan[1], K.B. Oganesyan[1 *], A.H. Gevorgyan[2], E.A. Ayryan[3], Yu.V. Rostovtsev[4]

[1] A.I. Alikhanyan National Science Lab, Yerevan Physics Institute, Alikhanian Brothers 2, 375036 Yerevan, Armenia

[2] Yerevan State University, Yerevan, Armenia

[3] Joint Institute for Nuclear Research, LIT, Dubna, Russia

[4] University of North Texas, Denton, TX, USA

[*] bsk@yerphi.am

## Abstract

The terahertz FEL ferromagnetic undulator has been created. The length of the undulator period is 9cm, number of the periods is 27. By means of selection and redistribution of magnetic elements the spread of amplitudes of the magnetic field was reduced to 7%. Additional windings of magnetic elements were used to compensate for residual spread. Required focusing gradient of magnetic field was obtained as a result of relative displacement of alternating poles with opposite charge of magnetic field along the x axis. Parameters of undulator including focusing properties in the horizontal plane were investigated.

## I. INTRODUCTION

There is increasingly interest in the development of Free electron lasers [1-73 and references therein]. Free-electron lasers (FELs) are composite systems of accelerators, electron beam (e-beam) optics, and undulators that produce widely tunable light with exceptional brightness at wavelengths down to hard x-rays for a broad range of studies. Effective energy exchange between the electron beam moving in an undulator and electromagnetic wave happens when resonance condition takes place. In this case electromagnetic wave advances electron beam by one radiation wavelength while electron beam passes one undulator period. When amplification process enters nonlinear stage, the energy losses by electrons become to be pronouncing which leads to the violation of the resonance condition and to the saturation of the amplification process.

Depending on the manner by which magnetic field is created two types of undulators can be distinguished: electromagnetic (including superconducting) and undulators on the permanent magnets. In [2] results of numerical computations carried out using three-dimensional magnetostatic code RADIA for different undulator schemes are presented. The maximum achievable magnetic field as the function of relations of the gap to the undulator period was found for each type of undulators. Currently undulators on the permanent rare-earth magnets are of the most interest due to their capacity to reach highest possible magnetic field as well as simplicity and compactness of their structure. But these magnets are very expensive, intractable and they have considerable spread of parameters. Here we present description of the undulator with cheaper ferrite magnets.

## II. CONSTRUCTION OF THE UNDULATOR

The highest value of undulator gain k is desirable in the magnetic undulators.

$$k = \frac{eB_0\lambda_u}{2\pi mc^2} = 0.943\,\lambda_u(cm)\times B_0(T), \tag{1}$$

where $\lambda_u$ - length of the period of undulator, $B_0$ - amplitude of magnetic field of undulator. In classical undulator radiation $k \sim 1$, the selection of magnetic field magnitude depends on $\lambda_u$ value. If undulator period is short, either high values of magnetic field are required, or electromagnets or permanent magnet from rare-earth elements are applied.

In undulators built for FEL in terahertz domain the length of the undulator period is about 10*cm*, therefore, magnetic field of about 1000 *Oe* is required. This allows usage of the cheapest and widespread ferrite metals magnets. Permanent magnets of 22220 type were used. The length of the undulator period is 9*cm*, number of the periods is 27. The width of magnets is 6*cm*, height - 2*cm*.

Preliminary calculations showed that 2.7 cm magnet length is sufficient to obtain required amplitude of magnetic field of undulator ( not less, than 0.1*T*). Unfortunately, when this size of magnet plates is used the distribution curve of magnetic field is far from sinusoidal. More comprehensive calculations by FEMLAB-3.1 program showed that in order to obtain sinusoidal distribution of magnetic field the length of magnets should be 3.2cm. The distribution of magnetic field for the duration of one period is presented in the Fig.1.

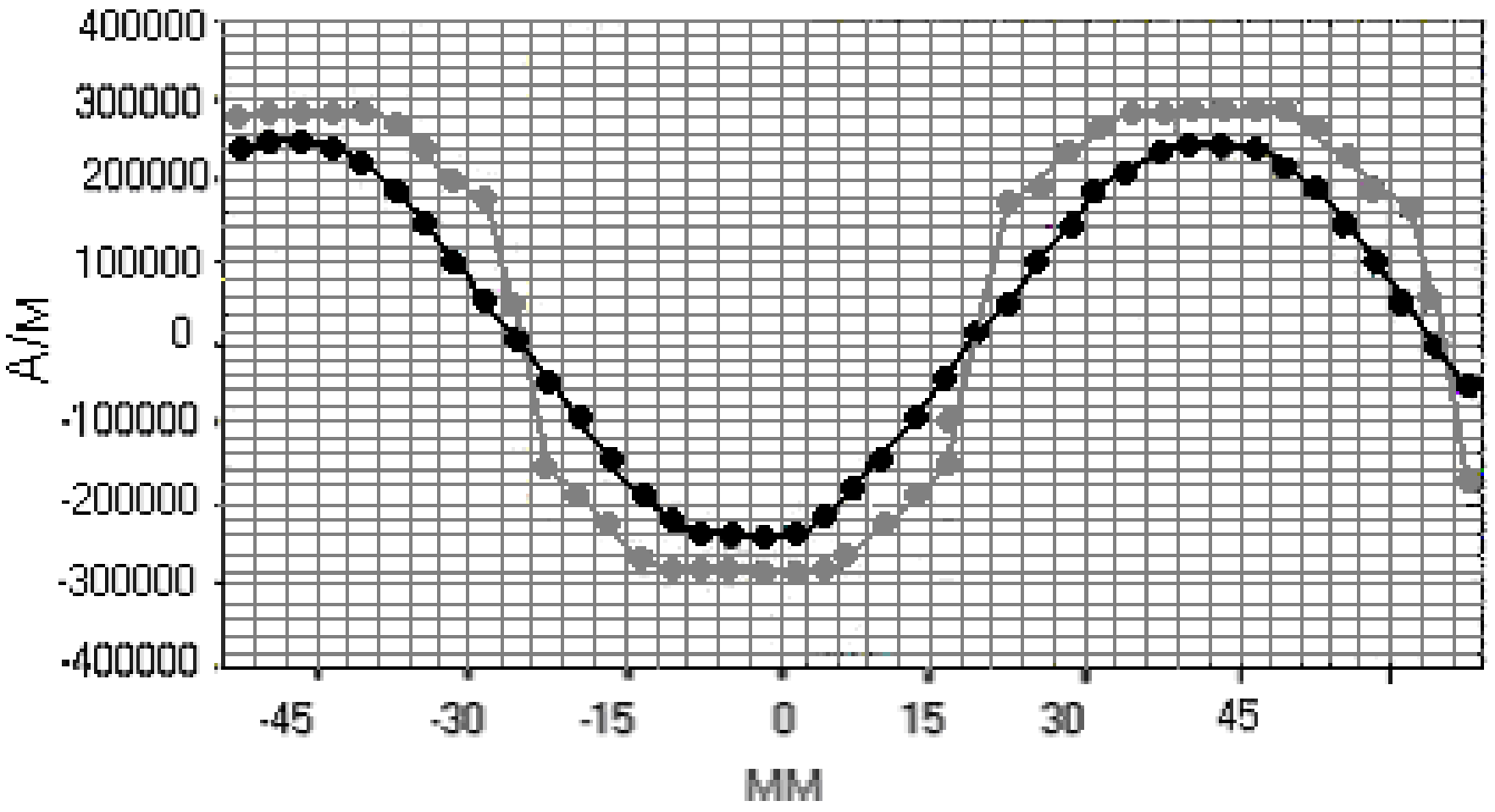

Fig. 1. The distribution of magnetic field for the duration of one period (black line - along the centre of undulator, gray line - close to poles).

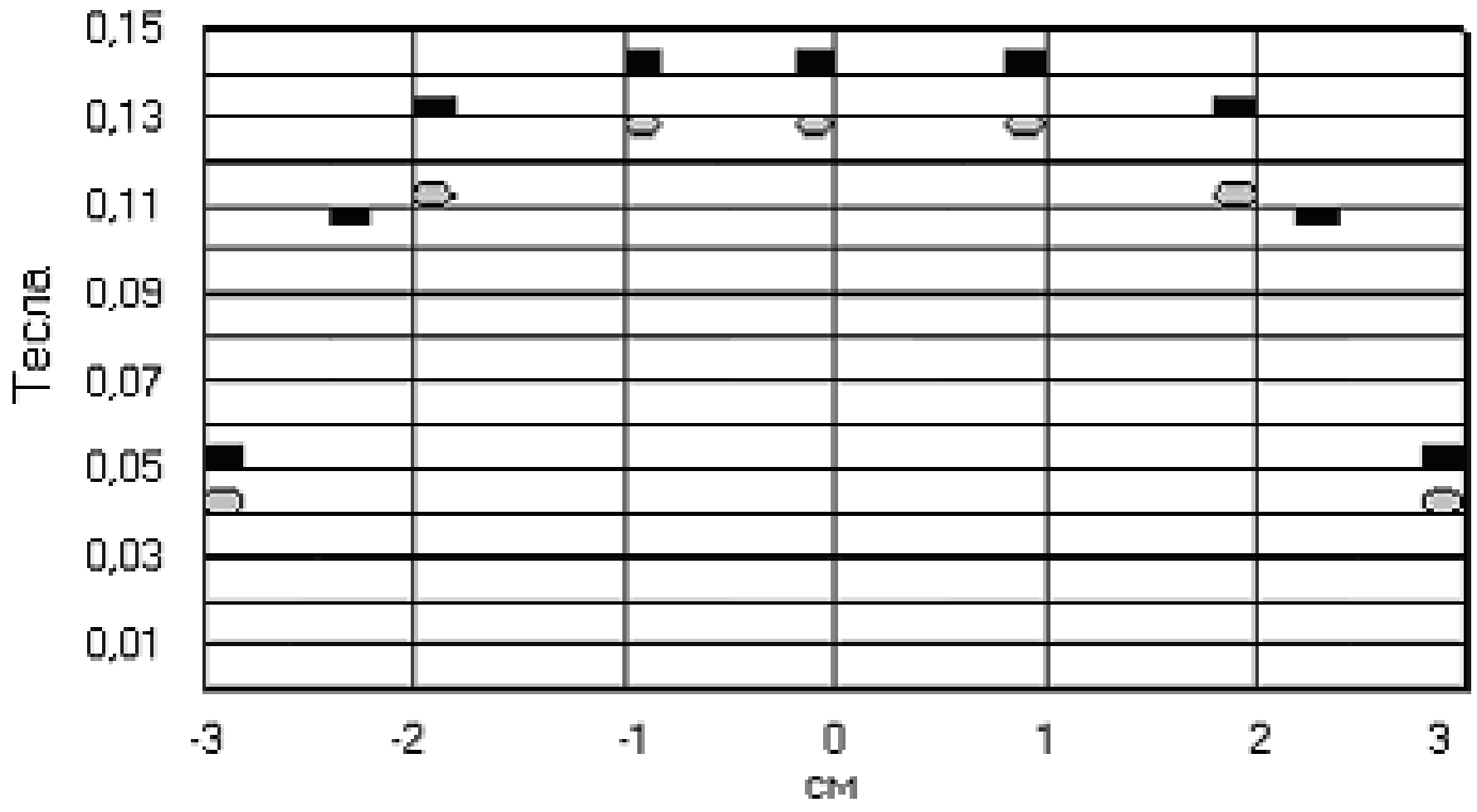

Fig 2. Horizontal distribution of magnetic field of undulator.
(Circles – results of measurements, squares – data point calculated using FEMLAB)

## III. CATEGORIZATION OF MAGNETS

The main problem while creating undulator was to reach the equal amplitude of magnetic field in all periods of undulator. One of the main disadvantages of undulator on permanent magnets is big spread of magnetic induction in different magnetic elements. The analysis of the spread of magnetic induction for all elements showed that initial spread of parameters was 40%, thus further utilization of such elements was impossible without special measures.

In order to decrease spread of parameters only magnetic elements with spread of magnetic induction around 18% were selected; therefore 20% of original magnetic elements were left out. In order to equate the amplitude of magnetic field of all periods two distinct methods were applied. Firstly, by means of selection and redistribution of magnetic elements we were able to reduce the spread of amplitudes of magnetic field to 7%. Secondly, additional windings of magnetic elements were used for the indemnification of residual spread. Conducted measurements showed that using relatively small winding with the current density equal to $3A/mm^2$ it is possible to reach sufficient compensation ($\sim 3\%$).

## IV. BEAM FOCUSING IN UNDULATOR

Widely used at undulators create natural focusing of electron beam only in vertical direction *y*. To generate electromagnetic radiation in millimeter range of wave-length visibly divergent low energy electron beams are used. Therefore e_ective focusing in both cross directions (*x* and *y*) is essential. According to work [3] beam uctuation wave magnitude in all three directions *x; y* and *z* meet the criterion:

$$k_x^2 + k_y^2 = k_u^2 \tag{2}$$

where

$$k_x = \sqrt{\frac{eq}{\gamma mc^2}}, \qquad k_y = \frac{2\pi}{\lambda_u}, \tag{3}$$

betatron wave number of quadrupole focusing in the horizontal plane (*q* - the force of quadrupole focusing). Equation (2) demonstrates that by adding focusing in the horizontal plane we decrease vertical (natural) focusing to the same extant. Generally equal focusing is reachable when the requirement is fulfilled. So the gradient of the feld in the horizontal direction equals:

$$q = \frac{k_x^2 \gamma mc^2}{e} = 272G/cm. \tag{4}$$

Different authors created schemes of undulators, with magnetic fields that are capable of focusing electron beam. The focusing gradient of magnetic field B is obtained through relative displacement of alternating poles with an opposite charge of a magnetic field along the *x* axis. Along the edges of magnetic gap predominance of magnetic field amplitudes of identical charge arises and consequently the necessary profile *By*(*x*) is created. The results of measurements and calculations conducted on distribution of magnetic field of undulator in horizontal direction are presented in Fig 2. These results demonstrate that linear area with gradient of the order of 300*G/cm* is about 1*cm* with the center displaced from magnetic element center by 2*cm*, and the field in the center of selected area decreases by 20%. In this case magnetic elements are placed off-centre of undulator by 2*cm*.

## V. MEASUREMENTS OF THE PARAMETERS OF UNDULATOR

Apart from distribution of magnetic field for tuning of magnetic elements of undulator it is essential to know the first and the second integrals of magnetic field along the undulator axis that characterize the angle and coordinate x of undulator beam output [4]. In [4] wire impulse method for rapid measurements of the undulator magnetic fields was proposed to be employ. This method was applied in some other works [5], [6]. The analogous device to measure magnetic characteristics of undulator was created by us. The scheme of this device is presented in the Fig.3.

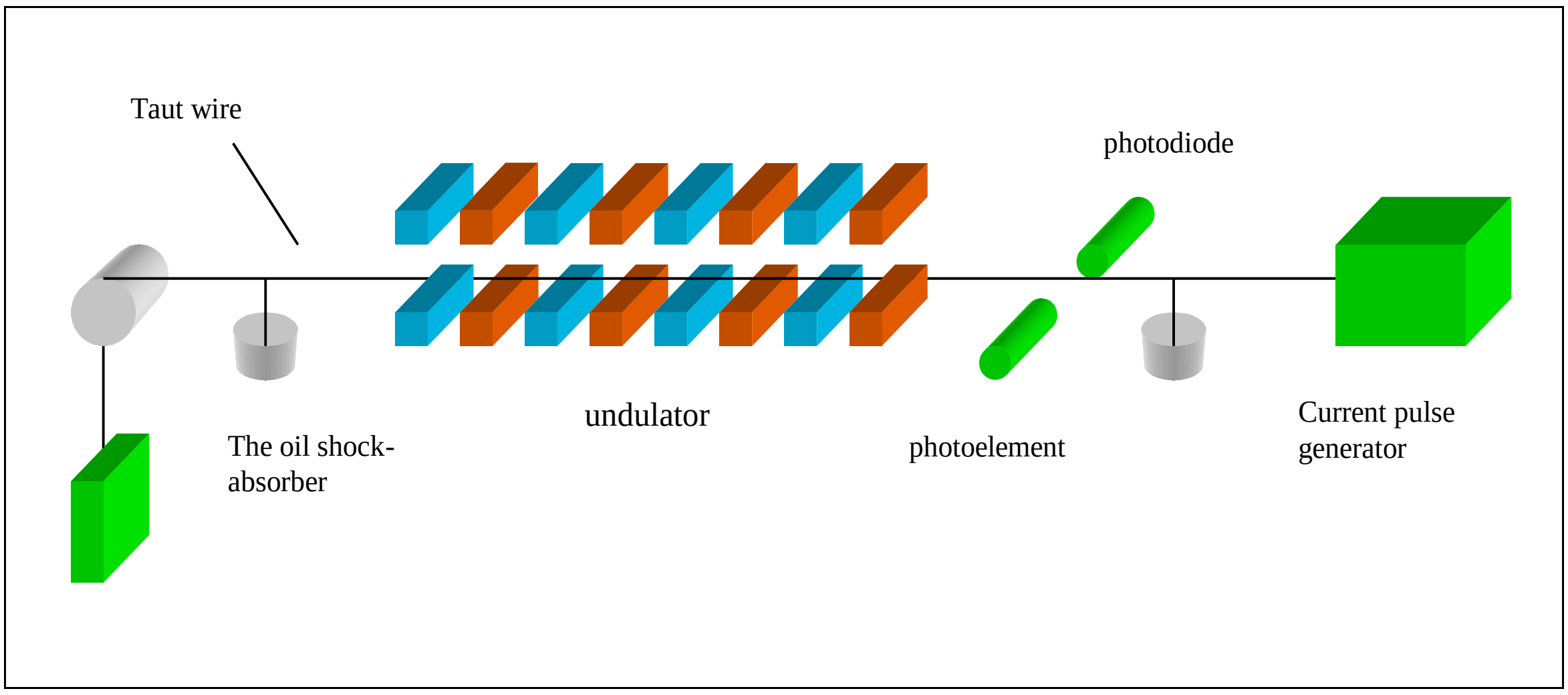

Fig.3. Installation diagram for measuring the integral characteristics of magnetic field of undulator.

The primary element of this device is wire, strained along the undulator axis. One side of the wire is fixed to the immovable support, and the other side thrown over a roller and stretched with the bob. Current impulse I is passed through the wire. In the wire the bump area is formed spreading along the wire in forward and backward directions at acoustic velocity:

$$v = \sqrt{T / P} \sim m / s, \qquad (5)$$

where $T$ is tighting force of thread, $P$ is the linear titer of thread material. The pickup signal at the wire bending receiver is proportional to the first integral of magnetic field (if short current impulse is used) or to the second integral (if long current impulse is used). In case of short current impulse the vibrational amplitude of the wire is as described in [6]:

$$A_1 = \frac{I_{01}\Delta t B_0 \lambda_u}{4\pi P v}. \qquad (6)$$

here $I_{01}$ - magnitude of the current in the wire, $\Delta t$ - duration of current impulse, $P$ linear density of the wire material. To observe the second integral of magnetic field the longer current impulse is required. During the duration of this impulse the acoustic wave has to pass whole undulator, thus following conditions should be fulfilled:

$$\Delta t v > L_{und}, \quad i.e. \quad \Delta t > (L_{und} / v) \qquad (7)$$

When working with 270*cm* long undulator (acoustic wave velocity is 300*ms*) current pulse duration is 10*ms*. The vibrational amplitude of the wire in case of long current impulse is:

$$A_2 = \frac{I_{02} B_0 \lambda_u}{48\pi^2 T}. \qquad (8)$$

where $T$- is tighting force of thread. In particularly, vibrational amplitude equals several mm when amplitude of current impulse is 1:5*A*.

In order to evaluate focusing properties of undulator in horizontal plane the transversal direction of the wire with current was investigated. The results are presented in Fig.4. Data in Fig. 4 shows that there is fairly well established focusing in horizontal plane.

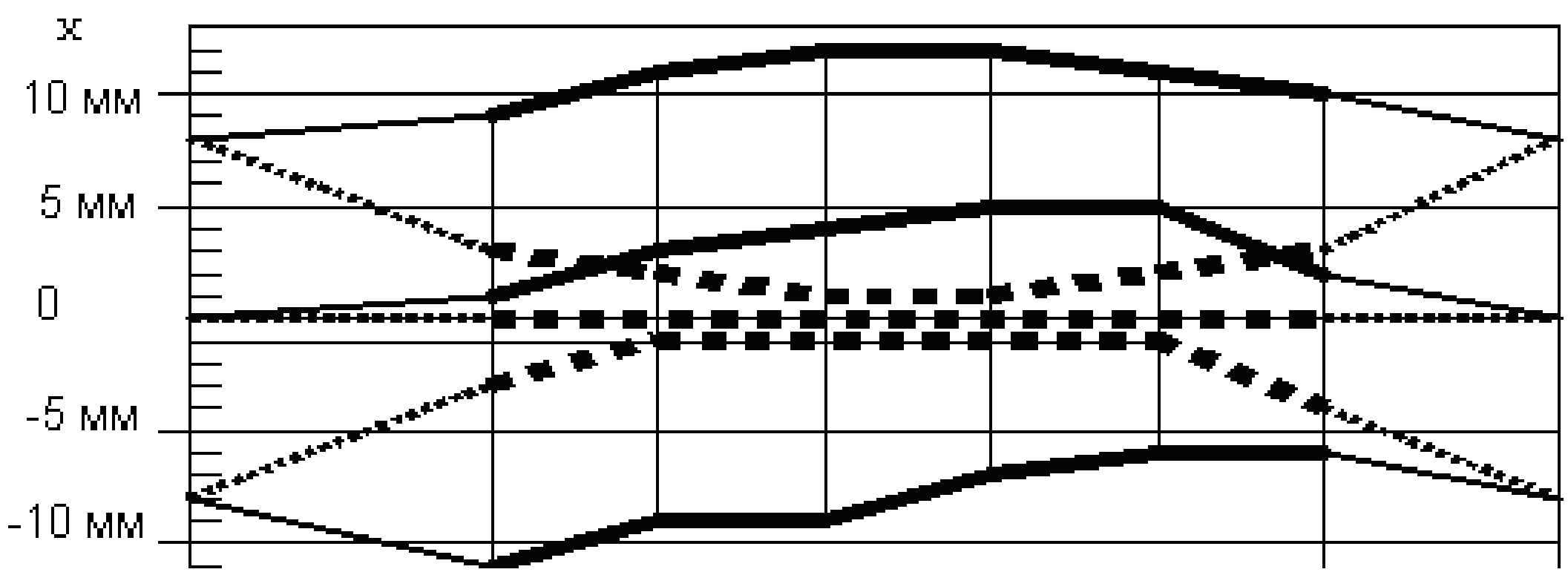

Fig 4. Transversal deflection of wire with current in undulator. Continuous line- magnetic elements placed abreast. Dotted line- magnets are displaced by 1 cm from central line.